\documentclass[10pt,amsmath,amssymb,twocolumn,superscriptaddress,showpacs,a4paper,floatfix,prb,aps]{revtex4-1}

\usepackage{geometry}                % See geometry.pdf to learn the layout options. There are lots.
\usepackage[parfill]{parskip}    % Activate to begin paragraphs with an empty line rather than an indent
\usepackage{graphicx}
\usepackage{array}
\usepackage{subfigure}
\usepackage{amsmath}
\usepackage{epstopdf}
\usepackage{stackengine}
\usepackage{placeins}
\usepackage{hhline}

\begin{document}
\title{2D-3D crossover in a dense electron liquid in silicon}

\author{Guy Matmon}
\email{g.matmon@ucl.ac.uk}
\affiliation{London Centre for Nanotechnology, University College London, 17-19 Gordon Street, London WC1H 0AH, United Kingdom}
\author{Eran Ginossar}
\affiliation{Advanced Technology Institute, University of Surrey, Guildford GU2 7XH, United Kingdom}
\author{Byron~J. Villis}
\author{Alex K\"olker}
\author{Tingbin Lim}
\author{Hari Solanki}
\author{Steven~R. Schofield}
\author{Neil~J. Curson}
\affiliation{London Centre for Nanotechnology, University College London, 17-19 Gordon Street, London WC1H 0AH, United Kingdom}
\author{Juerong Li}
\author{Ben~N. Murdin}
\affiliation{Advanced Technology Institute, University of Surrey, Guildford GU2 7XH, United Kingdom}
\author{Andrew~J. Fisher}
\affiliation{London Centre for Nanotechnology, University College London, 17-19 Gordon Street, London WC1H 0AH, United Kingdom}
\author{Gabriel Aeppli}
\affiliation{Institute of Physics, Ecole Polytechnique F\'ed\'erale de Lausanne (EPFL), 1015 Lausanne, Department of Physics, ETH Z\"urich, 8093 Z\"urich, and Paul Scherrer Institute, 5232 Villigen, Switzerland}

%\date{\today}                                           % Activate to display a given date or no date

\begin{abstract}
Doping of silicon via phosphine exposures alternating with molecular beam epitaxy overgrowth is a path to Si:P substrates for conventional microelectronics and quantum information technologies. The technique also provides a new and well-controlled material for systematic studies of two-dimensional lattices with a half-filled band. We show here that for a dense ($n_s=2.8\times 10^{14}$\,cm$^{-2}$) disordered two-dimensional array of P atoms, the full field angle-dependent magnetostransport is remarkably well described by classic weak localization theory with no corrections due to interaction effects. The two- to three-dimensional cross-over seen upon warming can also be interpreted using scaling concepts, developed for anistropic three-dimensional materials, which work remarkably except when the applied fields are nearly parallel to the conducting planes.
\end{abstract}

\maketitle

\section{Introduction}

The possibility of using single phosphorus (P) atoms in silicon as qubit hosts has driven the development of a fabrication technology for deterministic placement of the dopants which is very different from ordinary ion implantation/annealing paradigms. Hydrogen-passivated surfaces of Si(100) wafers are edited using a scanning tunneling microscope tip which removes protons, leaving behind exposed regions of silicon which can selectively bind phosphine (PH$_3$) molecules to incorporate P in the semiconductor. Subsequent annealing and silicon deposition then yield encapsulated two-dimensional P structures; iteration of all steps in this process can provide three-dimensional devices\cite{Schofield:2003fk,Scappucci:2015fk}.

Important components of any integrated electronic device technology are highly conducting sheets which can serve as current vias or capacitance/shielding plates, and this is accomplished by simply exposing large unpassivated areas of Si to phosphine. The result is very heavily doped two-dimensional silicon, with typically 1/180 P atoms/Si atom. For comparison, the solubility limit of P in bulk CZ-grown Si is 1/450 -- 1/250 impurities/Si atoms\cite{Chiou:2000fk,McKibbin:2014ng}, and the highest densities for activated subsurface ion implanted layers correspond to 1/80 P/Si\cite{Solmi:1996fv}. Furthermore, for the two-dimensional electron gases produced in Si MOSFETs the carrier densities are typically 10$^{12}$--10$^{13}$\,cm$^{-2}$, corresponding to 1/5$\cdot10^3$--1/5$\cdot10^4$ electrons/Si, where we assume an inversion layer thickness of 10\,nm. Typical parameters for some 2D systems appear in Table \ref{Tab:2D systems} and it is apparent the $\delta$-layer technique produces the densest 2D systems while retaining the coherence characteristics.

\begin{table}[htp]
\footnotesize
\caption{Typical growth and transport parameters for different 2D systems. $L$ and $L_\varphi$ are the elastic and inelastic scattering lengths, respectively. $n_s$ and $n$ are sheet and bulk electron densities, respectively, and $d$ is the layer thickness.}
\begin{center}
\begin{tabular}{lrrlll}
	\hhline{======}\\
	2D system&$L$&$L_{\varphi}$&$n_s$$\times10^{14}$&$n$$\times10^{21}$&$d$\\
	&[nm]&[nm]&[cm$^{-2}$]&[cm$^{-3}$]&[nm]\\
	\hline
	Ge:P $\delta$-layer\cite{Scappucci:2015fk}&5&184&0.63&\hspace{7pt}0.45&\hspace{5pt}1.49\\
	Current work&15&320&2.8&\hspace{7pt}0.28&10\\
	Si MOSFET\cite{Wheeler:1981fk}&48&252&0.07&\hspace{7pt}0.007&10\\
	QW 2DHG\cite{Simmons:2001td}&26&61&4.5$\cdot$10$^{-4}$&\hspace{7pt}2.25$\cdot$10$^{-5}$&20\\
	Graphene\cite{Lara-Avila:2011zl}&50&900&0.005&$\sim$0.005&$\sim$1\\
	\hhline{======}
\end{tabular}
\end{center}
\label{Tab:2D systems}
\end{table}

Beyond their potential for conventional and quantum electronics, the P $\delta$-layers are of fundamental physical interest because they realise a disordered two-dimensional Hubbard model where the sites are provided by the P dopants. The unprecedented high density allows access to what should be the simple (disordered) Fermi liquid regime for large t/U where t is the hopping integral and U is the on-site Coulomb interaction, where classical weak-localization (WL) theory should become exact.  

We have consequently set out to test this hypothesis, taking advantage not only of high quality $\delta$-layers of phosphorus in silicon, but also of a modern vector magnet which allows fields to be applied in arbitrary directions under software control. The density of dopants in the conducting slab in our sample corresponds to a typical interimpurity spacing of 1.5\,nm which is approximately the Bohr radius for P in Si. Theoretical estimates\cite{Saraiva:2015fk,Le:2017eh} of the Hubbard model parameters for Si:P show for inter-impurity spacing of 0.75\,nm the unscreened on-site energy $U\approx 40$\,meV would be similar to the tunnelling rate $t$. Additionally some reduction of $U$ is expected due to screening and hence it is likely that the experimental regime that we study here is well described by the metallic regime of the Hubbard model\cite{Georges:1996dmft}. On this backdrop, the focus of this study is a full temperature, magnitude and angle-dependent magnetoconductance (MC) set of experiments showing unprecedentedly rigorous demonstration of the validity of WL theory in two dimensions, including the remarkably simple formula where the in-plane field component, due to the thickness of the finite $\delta$-layer, simply increases the inelastic scattering rate to be inserted into the Hikami-Larkin-Nagaoka WL formula\cite{Hikami:1980vn} form (which is valid in the limit of zero thickness and depends only on the perpendicular component):
\begin{equation}
	\frac{1}{\tau_\varphi} \rightarrow \frac{1}{\tau_{\mathrm{eff}}}=\frac{1}{\tau_\varphi}+\frac{1}{\tau_B}
	\label{Eq:TauB}
\end{equation}
where $\tau_\varphi$ is the inelastic scattering time and $\tau_B$ is the additional phase-breaking time due to $B_\parallel$. Mensz and Wheeler\cite{Mensz:1987fk} originally demonstrated the applicability of Equation \ref{Eq:TauB} for silicon MOSFETs in tilted fields, and Mathur and Baranger\cite{Mathur:2001fr} later added substantial theoretical justification.  

The form of the Hikami-Larkin-Nagaoka formula with $\tau_{\mathrm{eff}}$, where $B_\parallel$ and $B_\perp$ are not interchangeable, is qualitatively different from anisotropic three-dimensional metals where the field components are interchangeable. We have exploited the vector magnet to explore the growth of this difference on cooling, and have discovered another remarkably simple form which characterizes all data except those with B nearly parallel to the planes:

\begin{equation}
	\Delta\sigma^p=\Delta\sigma_{\parallel}^p+\Delta\sigma_{\perp}^p
	\label{Eq:$p$-mean}
\end{equation}

In the limit where the dephasing length is short compared to the magnetic length $\sqrt{\Phi_0 / B}$ (where $\Phi_0$ is the magnetic flux quantum), the power-mean $p$ approaches unity, meaning that conductances affected by scattering parallel and perpendicular to the plane simply add as for parallel resistor shunts. In the opposite limit, $p$ becomes very large, meaning that whichever conductance is larger, namely that due either to in-plane or out-of-plane scattering, dominates entirely. For isotropic, three-dimensional disordered metals, the power law $p$ is confined to vary between 1 and 4, i.e. values larger than 4 are not allowed.  

We also find a collapse of the tilted-field MC onto the perpendicular-field MC by using an angle-dependent anisotropic scaling. We associate $p$ with a power-law approximation of Hikami's WL expression, and the scaling parameter with the effective anisotopy of the system. Together, we show that they map the WL correction at arbitrary field angles and magnitudes for a given system.

\section{Device fabrication and experimental setup}

\begin{figure*}
%\begin{figure}[!ht]
\centering
	\hspace{-5 mm}
	\subfigure[]{
%		\stackinset{l}{-17pt}{t}{-23pt}{\includegraphics[scale=1]{Fig_1a_Top.eps}}{\includegraphics[scale=1]{Fig_1a_Bottom_SG.eps}}
		\includegraphics[scale=1]{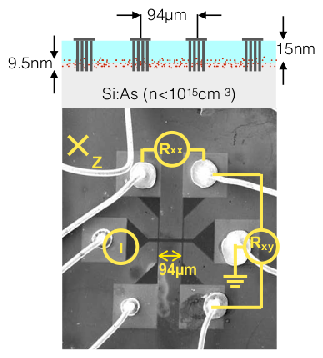}
		\label{Fig:Sample}
	\hspace{-5 mm}
	}
	\subfigure[]{
		\includegraphics[scale=1]{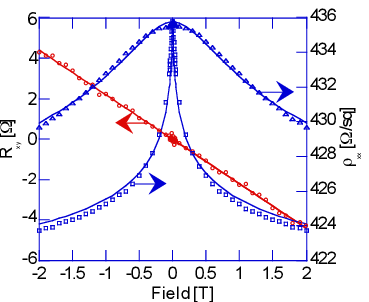}
		\label{Fig:Base Temp Fits}
	}
	\subfigure[]{
		\includegraphics[scale=1]{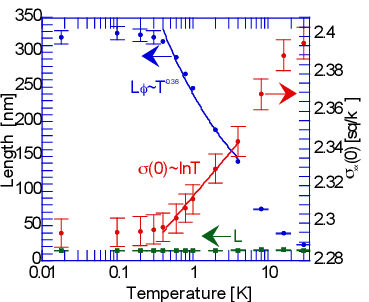}
		\label{Fig:MT Parameters}
	}
	\subfigure[]{
		\includegraphics[scale=1]{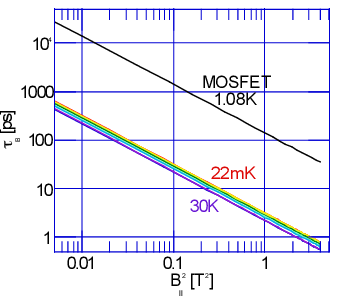}
		\label{Fig:TauB}
	}
	\caption{(Colour) Device and experiment. \ref{Fig:Sample} The Hall bar and the principal directions, and a schematic of the device cross section showing the arsenic-doped handle (light grey), the deposited phosphorus $\delta$-layer (red), the overgrown silicon (cyan) and the aluminum contacts. \ref{Fig:Base Temp Fits} MT curves at 22\,mK and their fits. Magnetoresistance (MR) with magnetic field out-of-plane in blue squares, in-plane in blue triangles, Hall resistance in red circles. \ref{Fig:MT Parameters} Temperature dependence of transport parameters. The blue circles are the phase coherence length. The unsaturated part which still has $L_\varphi \gg L$ shows a power-law dependence. The green squares are the mean free path, which is temperature-independent in this range. The red circles are conductivity at zero-field, showing logarithmic WL behavior. The error bars are 95\% confidence intervals. \ref{Fig:TauB} A temperature dependence of the experimental phase-breaking time $\tau_B$ vs. $B_\parallel^2$. The black line is MOSFET data from [\onlinecite{Mensz:1987fk}].}
	\label{Fig:Experimental}
%\end{figure}
\end{figure*}

Our test device was grown by saturation-dosing of PH$_3$ on a Si(100) flat surface, annealing for 2 minutes at 350\,C, then overgrowing 15\,nm of silicon at 250\,C without the use of a locking layer\cite{Hagmann:2018ul,Keizer:2015fk}. This is similar to the work by Goh \textit{et al.}\cite{Goh:2006ye}. The finished $\delta$-layer was processed into Hall bars, with mesa dimensions 94$\times$20\,$\mu \textrm{m}^2$. Scanning tunnelling microscopy (STM) measurements yielded a nominal sheet density of 2$\times$10$^{14}$\,cm$^{-2}$, and the measured Hall sheet density was 2.8$\times$10$^{14}$\,cm$^{-2}$ in the temperature range measured. The device was mounted in a dilution refrigerator with a 3D vector magnet capable of a field magnitude of 2\,T in all directions. The base temperature was $\sim$13\,mK (22\,mK with the field sweeping). Transverse and longitudinal resistance were measured for linear field sweeps, with the magnetic field out of plane ($Z$), in plane ($X$ and $Y$), and for angular sweeps at a field magnitude of 2\,T in all three planes. The temperature was varied from 22\,mK to 30\,K. Additionally, angular sweeps were performed at 22\,mK with different field magnitudes, from 2\,T to 0.1\,T. The current in all magnetoresistance (MR) measurements was 114\,nA. The current modulation frequency in all measurements was 7.6\,Hz. The main axes of the magnetic field were finely adjusted by applying a maximum in-plane field in the X and Y directions and minimizing the hall effect. An optical image and cross section of the sample appear in Figure \ref{Fig:Experimental}.

\section{Results}

Figure \ref{Fig:Base Temp Fits} shows magnetotransport (MT) curves with the field out of plane (transverse and longitudinal MR) and in plane (longitudinal MR). The fitted longitudinal MT is according to Hikami \textit{et al.} \cite{Hikami:1980vn} and Dugaev and Khemlnitskii \cite{Dugaev:1985fv}. The resulting layer thickness following Sullivan \textit{et al.}\cite{Sullivan:2004lr} is $9.44\pm0.47$\,nm, with a negligible temperature dependence. There was no measurable difference between the MR with the field in both in-plane directions. Figure \ref{Fig:MT Parameters} shows the temperature dependence of the transport parameters derived from the fits. We see that the conductivity scales as $\ln(T)$, as expected \cite{Altshuler:1980hh}. The phase coherence length $L_\varphi$ has a strong temperature dependence, and much like previous results in different material systems and thicknesses it saturates at low temperature\cite{Polley:2011ly,Cheung:1994mk,Fournier:2000qq,Ihn:2008hp,Ki:2008ye,Lin:2002rm}. We do not take this as an indication that the actual sample temperature saturates, since the magnitude of the zero-field peak caused by the superconducting transition of the aluminum contacts \cite{Polley:2011ly,Caplan:1965fk}  is still temperature-dependent at these low temperatures. The mean free path $L$ is only weakly temperature dependent, and as the temperature increases $L_\varphi$ tends towards $L$. We expect the WL approximation to break down as a result. The experimental values of $\tau_B$ are shown in Figure \ref{Fig:TauB} and we find that they are proportional to $B_\parallel^2$ as expected, with a weak temperature dependence\cite{Mensz:1987fk}. The proportionality coefficient is $1.5$ orders of magnitude lower than in previous MOSFET results, meaning that the in-plane field in our work has a larger phase-breaking effect. We attribute this to a rougher interface of the $\delta$-layer compared with the MOSFET\cite{Mensz:1987uq}.

\begin{figure*}
\centering
	\hspace{-1 cm}
	\subfigure[]{
		\includegraphics[scale=1.2]{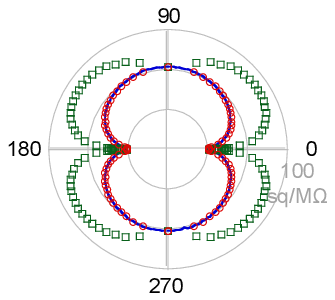}
		\label{Fig:Angular Lin Sum and p-Mean MC}
		}
	\hspace{-0.5 cm}
	\subfigure[]{
		\stackinset{l}{-1pt}{b}{-2pt}{\includegraphics[scale=1.2]{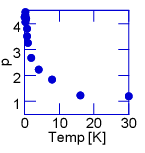}}{\includegraphics[scale=1.2]{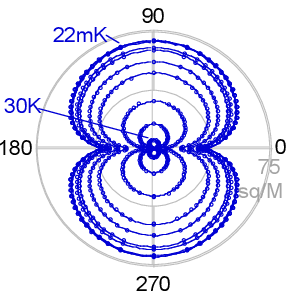}}
		\label{Fig:p-Mean Temp Dep}
		}
	\hspace{-0.35 cm}
	\subfigure[]{
		\stackinset{l}{0pt}{b}{-2pt}{\includegraphics[scale=1.2]{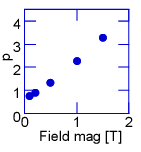}}{\includegraphics[scale=1.2]{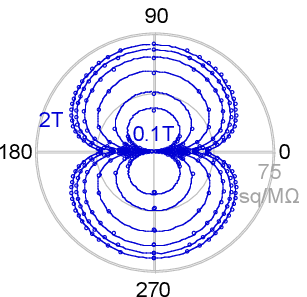}}
		\label{Fig:p-Mean Field Dep}
		}
	\hspace{-0.25 cm}
	\subfigure[]{
		\stackinset{l}{23pt}{t}{1pt}{\includegraphics[scale=1.0]{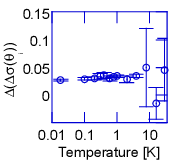}}{\includegraphics[scale=1.0]{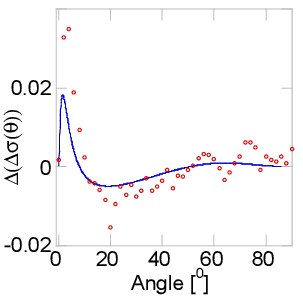}}
		\label{Fig:p-Mean Norm Dev Vs T}
		}		
	\caption{$p$-mean fits: \ref{Fig:Angular Lin Sum and p-Mean MC} Angular MC at a constant magnitude of 2\,T, 22\,mK. The blue line is the angle-dependent measurement. The green squares are a sum of the perpendicular and parallel field $\Delta\sigma$ values. The red circles are a $p$-mean of the perpendicular and parallel field $\Delta\sigma$ ($p=4.18\pm0.10$). \ref{Fig:p-Mean Temp Dep} Angular MC overlaid with $p$-mean fits. 2\,T field magnitude, different temperatures (from the outside at 22\,mK, 200\,mK, 400\,mK, 800\,mK, 2\,K, 4\,K, 8\,K, 16\,K, 30\,K). \ref{Fig:p-Mean Field Dep} 22\,mK, different field magnitudes (from the outside at 0.1\,T, 0.2\,T, 0.5\,T, 1.0\,T, 1.5\,T, 2.0\,T). Inset are the best values of p. \ref{Fig:p-Mean Norm Dev Vs T} Normalized deviation of the $p$-mean fit from the measured value of $\Delta\sigma$ (defined as $\Delta\left(\Delta\sigma(\theta)\right)\equiv\frac{\Delta\sigma(\theta)-\Delta\sigma_{\textrm{$p$-mean}}}{\Delta\sigma(\theta)}$) vs. field angle at 22\,mK showing that the most significant deviations are for small angles (i.e. nearly in-plane fields). The red symbols and the blue line are experimental data and calculation based on [\onlinecite{Mensz:1987fk}], respectively. Inset is $\Delta(\Delta\sigma)$ at the crossing angle $\theta_i$ (see Equations \ref{Eq:theta i} and \ref{Eq:crossing}) vs. temperature.}
	\label{Fig:p-Mean Dependences}
\end{figure*}

\begin{figure}
\centering
		{
		\includegraphics[scale=1]{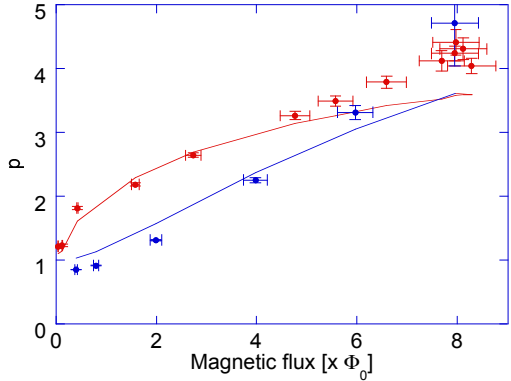}
		}
	\caption{(Colour) Values of $p$ vs the number of magnetic flux quanta through a circular closed trajectory with circumference $L_{\varphi}$. Values derived from temperature dependence are in red, and from field dependence in blue. The symbols are the fitted values of $p$ to measured results and the lines are calculated at $\theta_i$ (see Equation \ref{Eq:crossing}), with $\theta_i$ derived from the WL MC in [\onlinecite{Hikami:1980vn}] and [\onlinecite{Dugaev:1985fv}]. The X-axis errors originate in the $L_\varphi$ errors in Figure \ref{Fig:MT Parameters}.
 }
	\label{Fig:p Vs phi}
\end{figure}

Figure \ref{Fig:Angular Lin Sum and p-Mean MC} shows $\Delta\sigma_{xx}$ (hereafter referred to as $\Delta\sigma$) at 22\,mK and 2\,T as a function of the field angle in the ZX plane (blue). The sum of $\Delta\sigma$ caused by a parallel field and a perpendicular field at the appropriate magnitudes is shown in green. The traces intersect at 0$^{\circ}$, 90$^{\circ}$, 180$^{\circ}$ and 270$^{\circ}$ (measured from the plane), since only one field component is nonzero. At other angles this is clearly not the case. The red trace is a $p$-mean of the perpendicular and parallel field MC, where the fitted value of $p$ is $4.18\pm0.10$. The $p$-mean fits at a temperature range of 22\,mK -- 30\,K are shown in Figure \ref{Fig:p-Mean Temp Dep}; the quality of the fits remain good, and $p$ decreases as the temperature rises. $\Delta\sigma$ and $p$-mean fits at 22\,mK, for different field magnitudes, are shown in Figure \ref{Fig:p-Mean Field Dep}, and we see that $p$ increases with the magnitude of the field. Figure \ref{Fig:p-Mean Norm Dev Vs T} shows that the normalized deviation between the tilted-field measurement and the $p$-mean fit remains within 5\%. The symbols in Figure \ref{Fig:p Vs phi} are the experimental $p$-values, obtained from Figures \ref{Fig:p-Mean Temp Dep} and \ref{Fig:p-Mean Field Dep}, as a function of the number of magnetic flux quanta through the maximal area enclosed before dephasing, i.e. $A=\frac{L_{\varphi}^2}{4\pi}$. Separate sets of points are shown for measurements where the field is varied (blue) and where $L_\varphi$ was varied by changing the temperature (red). 

\section{Discussion}

To understand the origin of the $p$-mean we begin by considering the isotropic 3D case, where the MC depends only on the magnitude of $B$\cite{Kawabata:1980fk,Kawabata:1980kx,Yamanouchi:1967rm,Rosenbaum:1981fv}. At low and high fields it can be approximated by a local power law 
\begin{equation}
	\Delta\sigma(\alpha B)=\alpha^{2/p}\Delta\sigma(B).\\
	\label{Eq:power law}
\end{equation}
where $p=1,4$, respectively. For intermediate fields the relation also holds for $\alpha\approx 1$ and a $p$ which depends on the field $B$, so that the degree of non-linearity of the MC is encoded in $2/p(B)$. Let $\theta$ be the angle between the field and an arbitrarily chosen plane in an isotropic sample, so $B_\perp=B\sin \theta$ and $B_\parallel=B\cos\theta$; then by using Equation \ref{Eq:power law} the $p$-mean relation holds identically:

\begin{eqnarray}
	\left[\Delta\sigma(B_\perp)\right]^p&+&\left[\Delta\sigma(B_\parallel)\right]^p=\nonumber\\
	=(\cos \theta)^2\left[\Delta\sigma(B)\right]^p&+&(\sin \theta)^2\left[\Delta\sigma(B)\right]^p=\nonumber\\
	&=&\left[\Delta\sigma(B)\right]^p\nonumber\\
	\label{Eq:Isotropic p-mean}
\end{eqnarray}

A corresponding theory for anisotropic 3D samples, where $\Delta\sigma$ depends also on the angle $\theta$ of the field $B$ w.r.t. a reference plane in the sample, was discussed by Mauz \textit{et al.} \cite{Mauz:1997fb}. A one-parameter scaling rule for the magnetic field was deduced by which $\Delta\sigma$ for all angles collapses onto a single scaling curve:
\begin{equation}
	\Delta\sigma(B_\perp,B_\parallel)=f(B^*),\ \mbox{with}\ B^*=\frac{B}{\xi(\theta)}
	\label{Eq:conductivitycollapse}
\end{equation} 
where the anisotropy enters only through the effective field $B^*$, related to the physical field $B$ by the angle-dependent scale factor $\xi(\theta)$. $\xi(\theta)$ determines the effectiveness of a field at angle $\theta$ in disrupting WL relative to that of a perpendicular field.

\begin{figure}
\centering
	\subfigure[]{
		\includegraphics[scale=1]{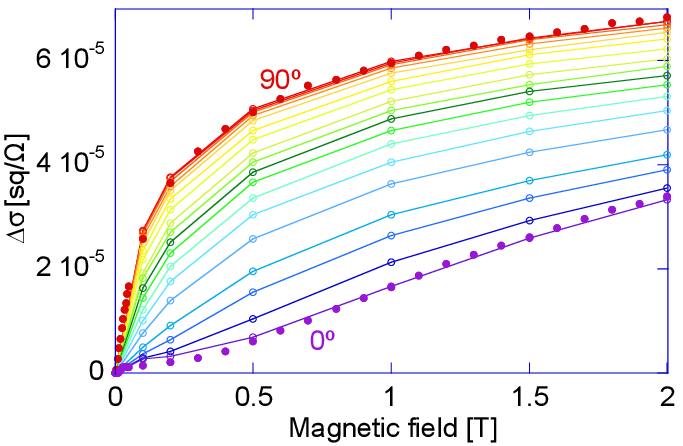}
		\label{Fig:Ds Vs Field Diff Angles Unscaled}
		}
	\hspace{-0.2cm}
	\subfigure[]{
		\stackinset{r}{3pt}{b}{21pt}{\includegraphics[scale=1]{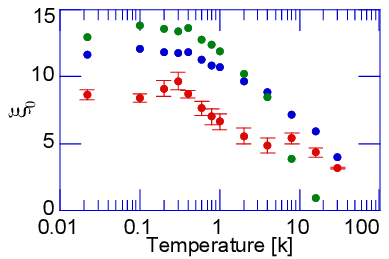}}{\includegraphics[scale=1]{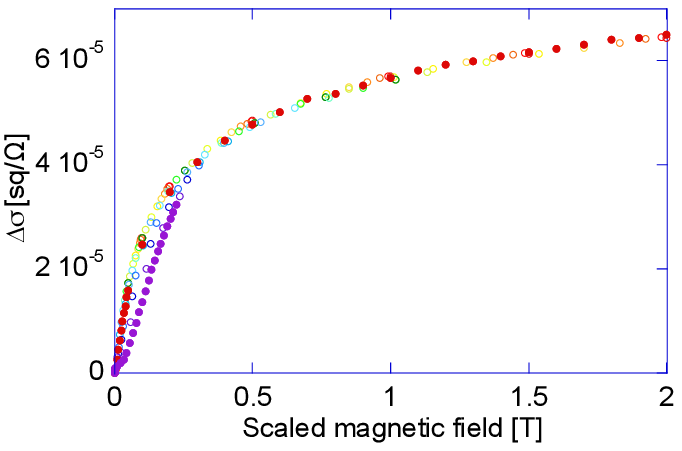}}
		\label{Fig:Ds Vs Field Diff Angles Scaled}
		}
	\caption{(Colour) Anisotropic field scaling: Figure \ref{Fig:Ds Vs Field Diff Angles Unscaled} shows $\Delta\sigma$ vs. field magnitude at different field angles, 22\,mK. Figure \ref{Fig:Ds Vs Field Diff Angles Scaled} shows same data as Figure \ref{Fig:Ds Vs Field Diff Angles Unscaled} but with the magnetic field scaled according to Equation \ref{Eq:scaling law} with $\xi_0=8.92$. The full red and violet circles are the $\Delta\sigma_\perp$ and $\Delta\sigma_\parallel$ values from Figure \ref{Fig:Base Temp Fits}, respectively. Inset in \ref{Fig:Ds Vs Field Diff Angles Scaled} are $\xi_0$ values at different temperatures, showing a stronger anisotropy at low temperatures. The red circles are best fits to the MC data, the blue and green ones are from Equations \ref{Eq:xi_0(theta)} and \ref{Eq:Asymptotic Approx}, respectively.}
	\label{Fig:Anisotropic Field Scaling}
\end{figure}

Following [\onlinecite{Mauz:1997fb}] we define a scaling for our single quasi-two-dimensional layer as
\begin{equation}
	\frac{1}{\xi(\theta)}=\sqrt{\frac{\cos^2(\theta)}{\xi_0^2}+\sin^2(\theta)}
	\label{Eq:scaling law}
\end{equation}
where $\xi_0\geq1$ denotes the effective anisotropy ratio, which can be approximately taken as constant over the angle and field ranges in the experiment (see Supplementary). We find a similar scaling collapse in our experimental data, as shown in Figure \ref{Fig:Anisotropic Field Scaling}: MC curves taken at a full range of angles $0^\circ\leq\theta\leq 90^\circ$ collapse onto the perpendicular field MC, which is well fitted by the expression derived in [\onlinecite{Hikami:1980vn}]. The temperature dependence of $\xi_0$ appears in the inset, showing that the system becomes more isotropic as the temperature increases.  We have also found that to a good approximation this same scaling collapse is empirically obeyed in Mathur's theory for a single layer with finite layer roughness and finite temperature \cite{Mathur:2001fr}. For large angles we see
\begin{eqnarray}
	\Delta\sigma^p&=&\left[f\left(\frac{B}{\xi(\theta)}\right)\right]^p\nonumber \approx\frac{f(B)^p}{\xi(\theta)^2}\nonumber\\
	&=&\left(\frac{\cos^2(\theta)}{\xi_0^2}+\sin^2(\theta)\right)f(B)^p
	\label{Eq:Scaled Power Law}
\end{eqnarray}
We see that if $2/p$ is taken to be the log-derivative of $f$ then
\begin{equation*}
	\sin^2(\theta)f\left(B\right)^p\approx f\left(B\sin(\theta)\right)^p.
\end{equation*}
Additionally, if $f(B)$ were a power-law (Eq.~\ref{Eq:power law}) then it would follow that
\begin{equation*}
	\frac{\cos^2(\theta)}{\xi_0^2}f\left(B\right)^p\approx f\left(B\frac{\cos(\theta)}{\xi_0}\right)^p.
\end{equation*}
In that case the p-mean relation (Equation~\ref{Eq:$p$-mean}) would hold as a result of scaling and power-law dependencies. Interestingly we find that even though $f(B)$ does not follow a power-law dependence, the empirical value of $p$ is modified such that the $p$-mean rule still applies and we can regard $2/p(B)$ as describing the interplay between $\Delta\sigma$ and the vector field \boldmath$B$ \unboldmath
in the 2D case.

We see that, provided a local power law exists, the $p$-mean is intrinsic to the isotropic bulk case. Furthermore, in the 2D case we successfully describe the tilted-field MC using the Hikami formula and a scaled field, and derive the $p$-mean from that scaling. We can now make several observations regarding the power law, $\xi_0$ and $p$. The ansatz in Equation \ref{Eq:power law} helped to demonstrate the $p$-mean relation for strictly field-magnitude related MC. We can show that for a particular field angle this power-law relation emerges directly from our scaling and $p$-mean observations. We define a `crossing angle' $\theta_i$ where the influence of the parallel and perpendicular fields, applied separately, is equal:  $\Delta\sigma_\perp(\theta_i)=\Delta\sigma_\parallel(\theta_i)$, where 

\begin{eqnarray}
	\Delta\sigma_\perp(\theta_i)&=\Delta\sigma(B\sin\theta_i,0)\nonumber\\
	\Delta\sigma_\parallel(\theta_i)&=\Delta\sigma(0,B\cos\theta_i) 	
	\label{Eq:theta i}
\end{eqnarray}
At this angle the $p$-mean exponent can be conveniently expressed as
\begin{equation}
	p=\left[\log_2 \left(\frac{\Delta\sigma(\theta_i)}{\Delta\sigma_{\perp}(\theta_i)}\right)\right]^{-1}
	\label{Eq:crossing}
\end{equation}
provided Equation \ref{Eq:$p$-mean} holds at the crossing point (which is empirically found to be the case, see Figure~\ref{Fig:p-Mean Norm Dev Vs T}). At the crossing angle we have, using Equation (\ref{Eq:conductivitycollapse}), 
\begin{eqnarray}
	f(B\sin\theta_i)&=&\Delta\sigma_\perp(\theta_i)=\Delta\sigma_\parallel(\theta_i)\nonumber\\
	&\approx&f\left(\frac{B\cos\theta_i}{\xi_0}\right)
\end{eqnarray}

and hence
\begin{equation}
	\cot\theta_i\approx\xi_0
	\label{Eq:xi_0(theta)}
\end{equation} 

i.e. the easily-measurable $\theta_i$ directly quantifies the MC anisotropy.

Using Equation \ref{Eq:scaling law} we see that $\xi(\theta_i)^{-1}=\sqrt{2}\sin\theta_i$. Once again using the empirical validity of Equation \ref{Eq:$p$-mean} at the crossing point, we see that

\begin{equation}
	p=\left[\log_2 \left( \frac{f(\frac{B}{\xi(\theta_i)})}{f(B\sin\theta_i)}\right) \right]^{-1}
	\label{Eq:p(theta)}
\end{equation} 

This expression for $p$ does not rely on the power-law assumption of Equation \ref{Eq:power law}. It then follows that

\begin{equation}
	f\left(\sqrt{2}B \sin\theta_i\right)=\left(\sqrt{2}\right)^{2/p}f(B \sin\theta_i)
\end{equation}
i.e. it emerges that $p$ encodes a power-law behavior of the MC scaling function at the crossing angle similar to Equation \ref{Eq:power law}.

$\xi_0$ can be related, at high fields, to $\tau_{el}$, $\tau_\varphi$ and $\tau_B$ (the elastic, inelastic and phase-breaking times, respectively) through an asymptotic approximation, i.e. at grazing angles (see Supplementary)
\begin{equation}
	\xi_0\approx\frac{4\pi D}{\Phi_0}B\tau_\mathrm{el}\left(\frac{e^{\psi(0.5)}}{2+\frac{\tau_\varphi}{\tau_B}}\frac{\tau_\varphi}{\tau_\mathrm{el}}-0.5\right)
	\label{Eq:Asymptotic Approx}
\end{equation}

where $D$ is the diffusion constant and $\psi$ is the digamma function. The values for Equation \ref{Eq:Asymptotic Approx} appear in the inset of Figure \ref{Fig:Ds Vs Field Diff Angles Scaled} (green circles) and we find the good agreement with the fitted data to be consistent with our understanding of $\xi_0$: a large anisotropy results in a $\theta_i$ that is close to the plane, making the asymptotic approximation in our calculation valid.

Equations \ref{Eq:xi_0(theta)} and \ref{Eq:p(theta)} give a full vector-characterization of anisotropic 2D-like WL. Given the results of a perpendicular- and parallel field MC measurement (e.g. the squares and triangles in Figure \ref{Fig:Base Temp Fits}, respectively), then the crossing angle $\theta_i$ can be found, for arbitrary magnetic fields, by interpolation. Both $\xi_0$ and $p(B)$ can then be calculated, and using Equations \ref{Eq:conductivitycollapse} and \ref{Eq:scaling law} or \ref{Eq:$p$-mean}, respectively, a full angle and field-magnitude map of $\Delta \sigma$ is constructed.

$\xi_0$ can also give an indication to the layer thickness; a natural interpretation of the scaling collapse in Equations \ref{Eq:conductivitycollapse} and \ref{Eq:scaling law} is that the MC is determined by the magnetic flux through an angle-dependent phase-breaking area. Taking the effective phase-breaking length perpendicular to the field as $L_\mathrm{eff}(\theta)=L_\varphi/\left(\xi(\theta)\pi\right)$, we might guess that the scaling approach would be valid until angles $\theta$ are reached such that the projection of $L_\mathrm{eff}(\theta)$ becomes comparable to the layer thickness, $d\approx L_\mathrm{eff}\cos\theta$.  Thus the breakdown of scaling gives a second -- independent -- estimate of $d$, which yields $d=13.7\,\mathrm{nm}$ at 22\,mK.

\section{Conclusions}
We have taken advantage of modern vector magnet and sample preparation technologies to revisit the problem of weak localization in silicon, where we have now implemented a two-dimensional half-filled disordered Hubbard model with t/U of order unity. To the best of our knowledge, the experiments represent the first attempt to examine the entire temperature and vector field dependence of the conductance for the realization of such a model, and we find astonishingly precise agreement with forty year-old theory, incorporating the effects of interface roughness but neglecting Coulomb interactions. The MC in a tilted field is found to be a $p$-mean of the MCs in the parallel and perpendicular field components separately. In analogy to the dependence of $\Delta\sigma$ on the magnitude of $B$ in the three-dimensional case of Kawabata\cite{Kawabata:1980fk}, we can regard $2/p$ as a generalised measure of the sensitivity of $\Delta\sigma$ to the varying $B$ vector in the two-dimensional case. 

There is a scaling collapse of the MC onto a single function of an angle-dependent effective magnetic field. Such a scaling form has previously been proposed for anisotropic 3D systems \cite{Mauz:1997fb}, but it is interesting that it holds also in the almost purely two-dimensional system formed by our $\delta$-layer.  Scaling is well obeyed for all field directions except when the field is very nearly in-plane. At these angles a crossover occurs to a different regime dominated by the parallel field; the out-of-plane confinement dominates the response to the field and the anisotropic 3D assumption breaks down.

\section*{Supplementary information:\\ Dependence of the scaling parameter $\xi_{0}$ on $\tau_{\varphi},\tau_{e}$}

Relying on the empirical validity of the scaling ansatz for a range
of fields and angles we can extract a dependence of the scaling parameter
$\xi_{0}$ on the system parameters. The scaling ansatz enables us
to obtain an expression for the vector magentoconductance $\Delta\sigma$
in terms of the perperndicular magnetoconductance $\Delta\sigma_{\perp}^{(s)}$
\cite{Hikami:1980vn} and the scaling function $\xi(\theta)$

\[
\Delta\sigma\sim\psi\left(\frac{1}{2}+\frac{B_{e}}{B_{\perp}}\right)-\psi\left(\frac{1}{2}+\frac{B_{\varphi}}{B_{\perp}}+\beta\frac{B^{2}\cos^{2}\theta}{B_{\perp}}\right)
\]

\[
\Delta\sigma_{\perp}^{(s)}\sim\psi\left(\frac{1}{2}+\frac{B_{e}}{B^*}\right)-\psi\left(\frac{1}{2}+\frac{B_{\varphi}}{B^*}\right)
\]
where $\beta$ represents the dephasing strength of the parallel magnetic
field, as has been discussed in several microscopic derivations \cite{Dugaev:1985fv,Mensz:1987fk,Mathur:2001fr} and $B^*=B/\xi(\theta)$ is the rescaled effective magnetic field. We can now try to compare the magentoconductances
assuming that $\theta\approx\theta_i$, an assumption which yielded a good description to $\xi_0$ as described in the main text with Eq. (11). For low temperatures we find that the crossing angle is relatively small. Equating $\Delta\sigma$ and $\Delta\sigma^{(s)}_{\perp}$ and numerically extracting the dependence of $\xi_0$ on temperature even for very small angles $\theta\approx 0.01$ yields a good approximation to the dependences depicted in the inset of Figure 4(b).

In order to obtain insight into the dependence of $\xi_0$ on the physical parameters we proceed with approximating each of the Digamma functions which appear above in the expressions for $\Delta\sigma$ and $\Delta\sigma^{(s)}_{\perp}$. For small angles and the values of $B,B_{\phi},B_e$ relevant for Figure 4(b), $\psi\left(\frac{1}{2}+\frac{B_{\varphi}}{B^*}\right)$ can be expanded close to $\frac{1}{2}$ i.e. where
\[
\psi\left(\frac{1}{2}+\epsilon\right)\approx\psi\left(\frac{1}{2}\right)+\frac{\pi^{2}}{2}\epsilon
\]
whereas the other Digamma functions are approximated by an asymptotically as logarithms since they possess large numerical arguments.
\[
\Delta\sigma\sim\log\left(\frac{B_{e}}{B_{\perp}}\right)-\log\left( \frac{B_{\varphi}}{B_{\perp}}+\beta\frac{B^{2}\cos^{2}\theta}{B_{\perp}}  \right)
\]

\[
\Delta\sigma_{\perp}^{(s)}\sim \log\left( \frac{1}{2}+\frac{B_e}{B}\xi_0\right)-\psi\left(\frac{1}{2}\right)
\]
where we find that it is sufficient to take the zero'th order approximation in the Taylor expansion for the last term.
By equating $\Delta\sigma$ and $\Delta\sigma^{(s)}_{\perp}$ and extracting $\xi_0$ we immediately obtain

\[
\xi_0\approx \frac{B}{B_e}\left( -\frac{1}{2}+\frac{B_e e^{\psi\left(\frac{1}{2}\right)}}{B_{\varphi}+\beta B^2}  \right)
\]
which can also be expressed in terms of the microscopic relaxation times as
\begin{equation}
\xi_0\approx\frac{4\pi D}{\Phi_0}B\tau_\mathrm{el}\left(\frac{e^{\psi(0.5)}}{2+\frac{\tau_\varphi}{\tau_B}}\frac{\tau_\varphi}{\tau_\mathrm{el}}-0.5\right)
\end{equation}

This expression reveals the dependence of the scaling parameter $\xi_0$ on the microscopic parameters and strengthens our assertion that the anisotropy of the system is highest at low temperature. This dependence is plotted in Figure 4(b) and while it works well to qualitatively describe the dependence for a large range of temperatures it is understandably less accurate for high temperature where the crossing angles are large.

We note that according to this derivation $\xi_{0}$ also observes
a dependence on the magnetic field. We argue that this manifests
the approximate nature of the scaling ansatz where $\xi_{0}$ is taken
constant. The constant $\xi_{0}$ is determined by a
fit and is mostly influenced by medium field strengths data. The reason
is that influence of $\xi_{0}$ is most important at medium range
fields because at small fields the magnetoconductance is small and
at large fields the magnetoconductance varies slowly with fields.
In practice this typical scaling factor, when applied to all measured
angles and fields performs well. We also see that the dependence of
$\xi_{0}$ with temperature, i.e. \textbf{$\tau_{\varphi}$ }is consistent
with trend that we find here.

\bibliographystyle{unsrt}
%\bibliography{p-Mean}

\end{document}